\documentclass[conference]{IEEEtran}
\IEEEoverridecommandlockouts

\usepackage{cite}
\usepackage{amsmath,amssymb,amsfonts}
\usepackage{algorithmic}
\usepackage{graphicx}
\usepackage{booktabs}
\usepackage{textcomp}
\usepackage{xcolor}
\usepackage{hyperref}
\usepackage{lipsum}
\def\BibTeX{{\rm B\kern-.05em{\sc i\kern-.025em b}\kern-.08em
    T\kern-.1667em\lower.7ex\hbox{E}\kern-.125emX}}

\usepackage{listings}
\usepackage{scalerel}
\usepackage{tikz}
\usetikzlibrary{svg.path}
\usepackage{svg}

\definecolor{orcidlogocol}{HTML}{A6CE39}
\tikzset{
  orcidlogo/.pic={
    \fill[orcidlogocol] svg{M256,128c0,70.7-57.3,128-128,128C57.3,256,0,198.7,0,128C0,57.3,57.3,0,128,0C198.7,0,256,57.3,256,128z};
    \fill[white] svg{M86.3,186.2H70.9V79.1h15.4v48.4V186.2z}
                 svg{M108.9,79.1h41.6c39.6,0,57,28.3,57,53.6c0,27.5-21.5,53.6-56.8,53.6h-41.8V79.1z M124.3,172.4h24.5c34.9,0,42.9-26.5,42.9-39.7c0-21.5-13.7-39.7-43.7-39.7h-23.7V172.4z}
                 svg{M88.7,56.8c0,5.5-4.5,10.1-10.1,10.1c-5.6,0-10.1-4.6-10.1-10.1c0-5.6,4.5-10.1,10.1-10.1C84.2,46.7,88.7,51.3,88.7,56.8z};
  }
}

\newcommand\orcidicon[1]{\href{https://orcid.org/#1}{\mbox{\scalerel*{
    \begin{tikzpicture}[yscale=-1,transform shape]
    \pic{orcidlogo};
    \end{tikzpicture}
}{|}}}}

\begin{document}

\title{Assessing LLMs for Front-end Software Architecture Knowledge\\
}

\author{
    \IEEEauthorblockN{Luiz Pedro Franciscatto Guerra}
    \IEEEauthorblockA{\textit{Computer Science,} 
    \textit{University of Victoria}\\
    Victoria, Canada \\
    luizguerra@uvic.ca \orcidicon{0009-0007-4020-9158}
}
\and
    \IEEEauthorblockN{Neil Ernst}
    \IEEEauthorblockA{\textit{Computer Science,} 
    \textit{University of Victoria}\\
    Victoria, Canada \\
    nernst@uvic.ca \orcidicon{0000-0001-5992-2366}
}
}


\maketitle

\begin{abstract}

Large Language Models (LLMs) have demonstrated significant promise in automating software development tasks, yet their capabilities with respect to software design tasks remains largely unclear.
This study investigates the capabilities of an LLM in understanding, reproducing, and generating structures within the complex VIPER architecture, a design pattern for iOS applications. 
We leverage Bloom's taxonomy to develop a comprehensive evaluation framework to assess the LLM's performance across different cognitive domains such as remembering, understanding, applying, analyzing, evaluating, and creating.
Experimental results, using \textit{ChatGPT 4 Turbo 2024-04-09}, reveal that the LLM excelled in higher-order tasks like evaluating and creating, but faced challenges with lower-order tasks requiring precise retrieval of architectural details.
These findings highlight both the potential of LLMs to reduce development costs and the barriers to their effective application in real-world software design scenarios.
This study proposes a benchmark format for assessing LLM capabilities in software architecture, aiming to contribute toward more robust and accessible AI-driven development tools.

\end{abstract}


\section{Introduction}

Specialized architectures like MVC (Model-View-Controller) offer rapid development advantages but can face scaling issues \cite{harun2019review}.
Advanced architectures such as VIPER (View-Interactor-Presenter-Entity-Router) provide a consistent structure for building software but demand significant learning and implementation time \cite{harun2019review, garcia2023viper}, posing challenges for new developers in communities favoring simpler approaches.

Large Language Models (LLMs) could provide assistance with these problems.
They are becoming excellent tools for developers (e.g. Microsoft's Copilot and Sourcegraph's Cody), excelling at generating repetitive code and documentation, but struggle with complex tasks \cite{liang2024large}.
Despite their growing utility, limited research explores how LLMs support front-end architectures like VIPER, as most studies focus on general project-level architecture or unrelated domains \cite{ardimento2024teaching, bariah2023large}.
This raises the question as to whether LLMs can comprehend project trees, identify underlying patterns, and assist in implementing new features while maintaining architectural consistency.

We examine the following research questions:
\begin{itemize}
    \item \textbf{RQ1}: To what extent do LLMs demonstrate understanding of fundamental front-end architectural patterns, using VIPER as a specific example?
    \item \textbf{RQ2}: Can LLMs accurately outline the structure of a VIPER module for that feature, specifying the roles and interactions of each component (View, Interactor, Presenter, Entity, Router), given a partial representation of the project of an iOS app (file tree)?
    \item \textbf{RQ3}: How does the quality of the solutions to architectural design challenges proposed by LLMs (in terms of adherence to VIPER principles, logical consistency, and scalability) compare to those provided by human iOS developers with varying expertise? 
    \item \textbf{RQ4}: How accurately can LLMs infer missing components, predict potential dependencies, or identify architectural inconsistencies?
\end{itemize}

To answer these questions:
    (a) We created an example iOS project. The project is written in the Swift programming language using the UIKit UI framework and implementing the VIPER architecture.
    (b) We created a set of questions using a knowledge taxonomy to ensure a broad coverage of questions.
    (c) We presented these questions to ChatGPT using its API.
    (d) We analyze and grade the AI's responses.
    Since a knowledge taxonomy was used, comparative analysis can then be used between the different cognitive levels.
    This paper benefits from the primary author's extensive experience with VIPER, gained from industry practice.

We selected the VIPER architecture \cite{garcia2023viper, gilbert_stoll_2014}, for being complex and incorporating a number of established patterns, being possible to generalize different responsibilities to other patterns (e.g. observer pattern instead of \textit{Presenter} tactic).

\section{Related Literature}

Recent advances in LLMs have shown promise in automating various software engineering tasks, but their application to front-end software architecture remains largely unexplored.

Ozkaya et al. \cite{ozkaya2023can} highlight the limited attention given to how design and architecture tasks can be effectively accomplished using generative AI-based software development tools. 
They emphasize the necessity of human expertise to assess not only the correctness of tool recommendations but also their fitness for purpose in code generation.
In the context of software engineering team automation, Lin \cite{lin2024llm} explores a scenario where each role in the software creation process is assumed by an LLM agent. 
However, the evaluation primarily focuses on the final generated code rather than the actual performance of the architect in the project.

Although the literature touches on various aspects of architecture and LLMs, direct applications to software architecture are still sparse.
Jahic and Hossain \cite{jahic2024state, hossain2024using} discuss system architecture, providing insights into how LLMs might be integrated into broader architectural frameworks.
Li \cite{li2024specllmexploringgenerationreview} explores the application of LLMs in chip architecture, suggesting potential parallels for software architecture.
Wei \cite{wei2024requirements} investigates the use of LLMs to generate software requirements, which indirectly influences architecture, particularly when following clean architecture principles. 
Several studies focus on the use of LLMs in generating UML diagrams \cite{ardimento2024teaching, conrardy2024image, ferrari2024model, wang2024llms}, 
Rasheed \cite{rasheed2024codepori} mentions the architecture of generated code, though it is often referenced indirectly in prompts rather than being a primary focus.
Mzid \cite{mzid2024attention} examines the detection of design patterns using LLMs.

\subsection{VIPER Architecture}

We selected the VIPER architecture \cite{garcia2023viper, gilbert_stoll_2014}, for being complex and with comparatively little online documentation that an LLM could have been trained on.
VIPER is a backronym to its five responsibilities. 
It is composed of five layers: the
    \textit{View} displays data and captures user input, 
    \textit{Interactor} handles business logic, 
    \textit{Presenter} acts as a mediator, handling presentation logic, 
    \textit{Entity} represents the data modelling, and
    \textit{Router} manages the navigation between screens.

It incorporates a number of established patterns (e.g. delegate), which could be generalized and adapted to other patterns (e.g., the observer pattern being utilized in the \textit{Presenter} tactic).
Additionally, it is a complex architecture with a large learning curve, contains multiple responsibilities and is aimed at large projects, as it has multiple advantages over other more common architectures, having good performance, testability, and modifiability metrics \cite{harun2019review}.

\section{Assessing Architectural Knowledge}

Multiple knowledge models have been developed to understand how developers comprehend software.
These models share common elements, including ``telltale code snippets and conjectures about the programming goals, and activities that bring the developer closer to the complete mental model" \cite{fekete2020comprehensive}.
The programmer's knowledge base typically consists of "domain expertise; coding knowledge, beacons, and plans; syntactic and semantic knowledge; domain goals; rules of programming discourse; and schemas" \cite{fekete2020comprehensive, o2003software}.
However, these software comprehension models are usually based on how developers understand complex code or build knowledge around source code for a project, and given the novelty of LLMs, they are not built or tested against it.

\subsection{Defining Architectural Knowledge}

Architectural knowledge encompasses a broader scope than just code comprehension. According to Clements et al. and Ozkaya \cite{clements2002software, ozkaya2023can}, architectural knowledge can be considered the union of five factors:
    \textit{Architecture design} (overall structure of the software system);
    \textit{Design decisions} (choices made during the architecture design process);
    \textit{Assumptions} (attributes assumed to be true during the architecture design);
    \textit{Context} (environment where the system will be used, e.g. hardware, operating system, network infrastructure); and
    \textit{other factors} (such as cost of development and maintenance).

For simplicity, we are choosing to follow Ozkaya's characterization \cite{ozkaya2023can} for defining architectural knowledge and the process of capturing design decisions.
Capturing design decisions and its assumptions requires knowing the context and the trade-offs.
Architectural knowledge and the ability to make meaningful trade-offs are expert skills and imply the experience of seeing similar examples over different situations.
If we want to use generative AI tools to assist in system development beyond method or class-level implementation tasks and toy examples, we must embrace incorporating architectural knowledge into the process, implying using architectural patterns, tactics, and design constructs to direct code generation.

\subsection{Assessing Architectural Knowledge}

A key aspect of learning software design is conveying architectural knowledge \cite{hu2013nature}.
This involves understanding not only algorithms and data structures, but also higher-level concerns (e.g. system structure and component interactions).
Architectural design necessitates analyzing software requirements to capture functional and behavioral aspects, defining not only what a system is, but also what it does.
Architectural design requires analyzing requirements to define both the structure and behavior of a system.
Studying styles like layered architectures and Model-View-Controller provides a foundation for organizing complex systems, though learning design processes remains challenging \cite{hu2013nature}. 
Hu highlights that students often lack process knowledge—understanding the phases, paradigms, and methodologies essential for creating design artifacts.

Bloom's taxonomy provides a structured framework for assessing architecture knowledge by aligning learning objectives to benchmark approaches. 
It classifies cognitive learning into six hierarchical levels, from basic to complex \cite{krathwohl2002revision}:
    \textbf{Remembering}: recalling facts (e.g., listing software architectures).
    \textbf{Understanding}: demonstrating comprehension (e.g., explaining differences between architectures).
    \textbf{Application}: using information in new contexts (e.g., developing an app using MVC).
    \textbf{Analysis}: breaking down components (e.g., analyzing interactions in hybrid app architectures).
    \textbf{Evaluating}: making judgments based on criteria (e.g., assessing performance implications of a design pattern).
    \textbf{Creating}: combining elements to design novel solutions (e.g., crafting a new architecture for a specific problem).

Traditional approaches to teaching software architecture focus on textbooks and standardized solutions, as these methods were developed before the advent of LLMs. 
These tools impact information retrieval and collaboration, aiding or hindering developers depending on their usage \cite{haque2024information}.
While AI tools show efficiency gains, a new approach is needed to leverage LLMs’ capabilities in acquiring and applying architectural knowledge, as they have the potential to be transformative tools.

\section{Method}

To investigate the capabilities of the LLM in understanding and working within the VIPER architecture, we designed an investigation setup that covers various aspects of the architecture and its implementation in Swift.
We create architectural knowledge questions about the framework, and assess the LLM performance using Bloom's taxonomy \cite{krathwohl2002revision}.

\subsection{Project Structure and Language}

The selected project\footnote{\url{https://github.com/LuizGuerra/Frontend-with-LLM-Project}} is a small Swift project, using UIKit framework, and follows the VIPER architecture.
The code was built by the author, incorporating contributions from novice developers.
The project includes realistic architectural errors reflective of the team's learning process at the time.
It contains multiple VIPER modules, and each one of them holds its components: View, Interactor, Presenter, and Router.
Entities are an exception, since they are commonly shared between multiple features, being usually found in a shared folder.
    
Protocols, although not included in the VIPER nomenclature, are a common design pattern used together with VIPER on Swift, commonly referred to as the Delegate Design Pattern---therefore the LLM will also be tested on this separation of concerns.

\subsection{Question Design}

This architecture assesses the LLM's ability to respond and generate insights across varying cognitive levels. 
A diverse question set, based on Bloom's taxonomy, helped to ensure a more comprehensive evaluation:

\begin{itemize} \item \textbf{Remembering}: Identify file functionalities within VIPER modules. \item \textbf{Understanding}: Explain file purposes in context (e.g., the role of Business Logic in the \textit{MovieHomeScreen} module). \item \textbf{Applying}: Describe data flows between VIPER elements and propose implementation strategies. \item \textbf{Analyzing}: Detect patterns, deviations, and bottlenecks in adherence to VIPER standards. \item \textbf{Evaluating}: Critique VIPER’s effectiveness, highlighting strengths, limitations, and areas for improvement. \item \textbf{Creating}: Suggest new features or refactor existing ones, considering localization, analytics, and development strategies. \end{itemize}

Questions are open-ended, allowing multiple approaches to be correct. Some intentionally lack definitive answers (e.g., data flows for non-existing features) to evaluate handling of ambiguity and hallucinations.

\begin{table*}[t]
    \centering
    \textbf{Performance by Knowledge Level}\par\medskip
    \begin{tabular}{l|c|c|c|c|c|c}
        \toprule
        \textbf{Knowledge Level} & \textbf{Remembering} & \textbf{Understanding} & \textbf{Applying} & \textbf{Analyzing} & \textbf{Evaluating} & \textbf{Creating} \\ \hline
        \textbf{Grade (\%)}             & 83\% & 94\%& 88\%& 100\%& 94\%& 100\%\\ 
        \textbf{\# Correct}             & 5 & 7 & 6 & 5  & 7  & 7 \\ 
        \textbf{\# Partially Correct}   & 0 & 1 & 2 & 0  & 1  & 0 \\ 
        \textbf{\# Incorrect}           & 1 & 0 & 0 & 0  & 0  & 0 \\ 
        \textbf{Total Questions}        & 6 & 8 & 8 & 5  & 8  & 7 \\ 
        \bottomrule
    \end{tabular}
    \caption{Model performance on question set. Grade for each level is calculated as the mean of all questions on said level, converted to percentage.}
    \label{tab:knowledge_levels}
\end{table*}

\subsection{Model and Queries}

The objective is to probe if the current state-of-the-art (SOTA) AI model is consistent.
At the time of running this study, the model was OpenAI's \textit{ChatGPT 4 Turbo 2024-04-09}.

LLM evaluations were conducted using the OpenAI API, with all prompts being structured as: 
\begin{lstlisting}
    Given the context that this is an iOS 
        project using VIPER architecture,
        answer the following question.
    <Query from Bloom's taxonomy>
    The project tree is the following:
    <Project Tree>
\end{lstlisting}

Note that no actual code is given to the model's context. 
Responses from the LLM were then collected and stored in a \textit{CSV} file for further analysis.

To ensure the evaluation accurately targeted the intended research questions, a preliminary run was taken with ChatGPT 3.5-Turbo. 
The goal was to eliminate any potential interference from unclear wording or grammatical issues.
In the case of misunderstandings, the question was iterated over until it was sufficiently clear.
The replies from the final model were registered and then graded.

\section{Preliminary Results}

For the grading, each answer could be equal to 1 point (correct), 0.5 points (partially correct), and 0 points (incorrect).
We considered an answer \textit{correct} when the model answered the question satisfactorily; 
\textit{Partially correct} when it made some mistakes in the answer, but it is not completely incorrect;
And \textit{incorrect} when it was not able to give the correct answer.

Each level of the taxonomy has between 4 and 7 questions, totaling 43 questions, and the final results can be seen in Table \ref{tab:knowledge_levels}. 
The grade for each level represents the mean score of all questions within that level, converted to a percentage.

Perhaps surprisingly, the model performed better at the later levels in Bloom's Taxonomy, scoring 100\% in \textit{Analyzing}, 94\% in \textit{Evaluating}, and 100\% in \textit{Creating}.
In the first levels, the model did not have bad scores, but it is comparatively lower, with 83\% in \textit{Remembering}, 94\% in \textit{Understanding}, and 88\% in \textit{Applying}.
This might demonstrate that the model is capable of creating actions and generating code, but fails with specific information retrieval without using code.

However, it is important to note that the combined ‘grade’ needs to be viewed with some skepticism due to the varying number of questions per knowledge level.
A simple average across the 'Grade (\%)' row would give equal weight to each level, despite the difference in number of questions. 

\section{Discussion} 

The LLM demonstrated a promising grasp of front-end architectural principles (RQ1), showing potential in high-level tasks like proposing features and critiquing architecture, though its software design understanding remains limited, suggesting architects will maintain a critical role.
Regarding outlining VIPER modules (RQ2), it shows capability in contributing to feature development and generating boilerplate code, suggesting structural understanding, yet struggled with describing component data flow.
The comparative quality of the LLMs architectural design solutions against human developers (RQ3) remains a subject for future investigation.
Finally, concerning its ability to infer missing components and dependencies (RQ4),
    its success in "Analyzing" suggests a capacity for identifying architectural inconsistencies;
    While constrained by lack of direct code access, necessitating higher-level suggestions, the LLM often correctly assumed relationships and proposed flexible solutions adaptable to existing project structure.

Integrating architectural knowledge into LLM training and using more complex datasets could enhance their ability to address real-world challenges.
Employing Bloom's taxonomy highlights the model's proficiency in creative and evaluative tasks but reveals gaps in detailed information retrieval and precise architectural application.

Improved LLM performance could lower development costs, support novice developers, and encourage robust design adoption, positioning LLMs as valuable tools rather than replacements in the software development process.

\subsection{Limitations}

While this study offers insights into LLM capabilities in understanding and applying software architecture concepts, particularly VIPER, several limitations must be acknowledged.

The evaluated project, developed by novices, contained architecture inconsistencies, providing a small-scale test case.
Larger, older projects with technical debt might yield different insights.
Using custom projects with designed errors could broaden the findings.

The study focused solely on VIPER in Swift, a niche architecture, which might limit generalization to other patterns like MVVM.
Additionally, evaluations relied on a single model, \textit{ChatGPT 4 Turbo 2024-04-09}, which may not reflect the behavior of other LLMs.
Future advancements or alternative models might alter the results.

The model excelled in higher-order tasks (e.g., creating and evaluating) but struggled with lower-order-tasks (e.g., remembering and applying), likely due to project inconsistencies or limited contextual data (e.g., only project trees, not full source code).
The "black-box" nature of LLMs also complicates understanding their decision-making, underscoring the need for interpretability tools and systematic error analysis.

Bloom's taxonomy provided a useful framework for assessing cognitive performance, but may not fully capture the complexity of software architecture tasks.
The questions were tuned for clarity, which can introduce bias, and the
manual response evaluation introduced subjectivity, though we will incorporate inter-rater reliability in future to monitor consistency.
Improved validation methods and developer feedback on LLM outputs will further refine these findings.

\subsection{Future Work}

We plan to expand our analysis to a larger project.
This shift necessitates revising the question set to align with the new repository.

To manage the increased complexity, we are incorporating inter-rater reliability into the question development process.
Questions will be independently created by a group of graduate students and subsequently evaluated anonymously by a separate panel of graduate peers.
Based on their feedback, questions will be refined, removed, or added until consensus is achieved.

We also intend to include an open-source LLM as a baseline and involve developers experienced with the architecture to evaluate the responses.
The finalized question set will be tested on the latest state-of-the-art model.

\section{Conclusion}

This study offers positive preliminary insights into LLM capabilities for architectural tasks, with potential for higher-order cognitive tasks, particularly in analyzing, evaluating and creating within the VIPER architectural pattern.
We achieve it by employing a methodology grounded in Bloom's Taxonomy and analyzing open-ended responses to architectural challenges.

It also establishes a structured approach for evaluating potential and limitations on LLMs cognitive abilities for Software Design and Software Engineering.
This framework provides a foundation for future research to more systematically explore the potential and limitations of LLMs in Software Engineering.

Despite these promising results, the study also reveals barriers to widespread application, including limited context information and occasional struggles with information retrieval.
As LLM technology evolves, improvements in training datasets, larger contexts, and interpretability could address these issues, enabling broader and more robust adoption.

\bibliographystyle{IEEEtran}
\bibliography{src}

\end{document}